\documentclass[aip,amsmath,amssymb,reprint]{revtex4-2}

\usepackage{graphicx}
\usepackage{dcolumn}
\usepackage{bm}

\usepackage[utf8]{inputenc}
\usepackage[T1]{fontenc}
\usepackage{mathptmx}
\usepackage{etoolbox}

\usepackage{booktabs}
\usepackage{soul}

\usepackage[caption=false, position=top]{subfig}
\usepackage{xcolor}

\makeatletter
\def\@email#1#2{%
 \endgroup
 \patchcmd{\titleblock@produce}
  {\frontmatter@RRAPformat}
  {\frontmatter@RRAPformat{\produce@RRAP{*#1\href{mailto:#2}{#2}}}\frontmatter@RRAPformat}
  {}{}
}%
\makeatother
\begin{document}

\preprint{AIP/123-QED}

\title[Sample title]{Spreading of graphene oxide suspensions droplets on smooth surfaces}

\author{J.A. Quirke}

\author{M.E. M\"{o}bius}
 \email{mobiusm@tcd.ie}
\affiliation{School of Physics, AMBER, CRANN, Trinity College Dublin, Dublin 2, Ireland}

\date{\today}

\begin{abstract}
Understanding and predicting the spreading of droplets on solid surfaces is crucial in many applications such as inkjet printing, printed electronics and spray coating where the fluid is a suspension and in general non-Newtonian. However, many models that predict the maximum spreading diameter usually only apply to Newtonian fluids. Here we study experimentally and theoretically the maximum spreading diameter of graphene oxide suspension droplets impacting on a smooth surface for a wide range of concentrations and impact velocities of up to $6$ g/l and $3$ m/s, respectively. As the particle concentration increases the rheological behaviour changes from a viscous fluid to a shear-thinning yield stress fluid and the maximum spreading diameter decreases. The rheology for all concentrations is well described by a Herschel-Bulkley model that allows us to determine the characteristic viscosity during spreading. We use this viscosity to develop an energy balance model that takes into account the viscous dissipation and change in surface energies to find the maximum spread diameter for a given impact velocity. The model contains one non-dimensional parameter that encodes both the dynamic contact angle during spreading and the droplet shape at maximum spread. Our model is in good agreement with our data at all concentrations and agrees well with literature data on Newtonian fluids. Furthermore, the model gives the correct limits in the viscous and capillary regime and can be solved analytically for Newtonian fluids. 

\end{abstract}

\maketitle

\section{\label{sec:intro}Introduction}

The precise control of droplet impacts on solid surfaces plays an important role in many industrial applications, from inkjet printing \cite{lohse2022}, gelled fuels\cite{Rahimi2011}, forensics \cite{Laan2014, Yokoyama2022} to printed electronics \cite{glasser2019, finn2014} where the resolution of the printed patterns depends on the initial droplet diameter and its spreading behaviour. As a liquid droplet with radius $R_0$ impacts on a smooth solid surface, it spreads to a maximum radius $R$ that depends on the impact velocity $U_0$, viscosity $\eta$, surface tension $\sigma$ and density $\rho$ of the fluid, along with the wetting properties \cite{lin2018,DeGoede2019,Yarin2006,Josserand2016}. Considerable work has been carried out to measure and predict the maximum spreading diameter for Newtonian fluids which is usually quantified by the spreading ratio $\beta_{max}=R/R_0$. Several semi-empirical scaling relations between $\beta_{max}$, the Reynolds number $Re=\rho (2R_0) U_0/\eta$ and Weber number $We=\rho (2R_0) U_0^2/\sigma$ have been established \cite{Scheller1995,Clanet2004,Roisman2009,Laan2014, Lee2015}. Another way to estimate the value of $\beta_{max}$ is through energy conservation models that consider the viscous dissipation and change in surface energy during the spreading process \cite{aksoy2022,Chandra1991,Pasandideh-Fard1996,Mao1997,Ukiwe2005,Wildeman2016,huang2018,du2021,worner2023, Zhao2018}. 
 
In the case of non-Newtonian, shear thinning fluids the shear rates and therefore the viscosity can decrease by orders of magnitudes during the impact and spreading process which needs to be accounted for in order to predict the maximum spreading diameter. While several studies have emerged in recent years addressing this question\cite{liu2024,luu2009, Balzan2021a, Piskunov2021, Rahimi2011, An2012a, Yokoyama2022, Dechelette2010, German2009a,gorin2022}, a complete description is still lacking and further experiments are required to understand the relevant time and energy scales that govern the spreading process\cite{shah2024}. We use suspensions of high-aspect ratio 2D nanoparticle suspensions to address this question. At low concentrations these type of suspensions are Newtonian\cite{barwich2015} and their spreading behaviour largely follows that of the solvent\cite{lee2014}. However, no studies thus far have examined the spreading behaviour for dense 2D nanoparticle suspensions where the fluid is shear thinning and develops a yield stress.  

Here, we study the spreading dynamics of aqueous graphene oxide (GO) suspensions droplets on a smooth glass slide. GO are high aspect ratio flakes consisting of a graphene hexagonal carbon lattice with its basal plane and edges containing oxygen based functional groups \cite{Du2020} and is therefore readily dispersible in water. The rheological behaviour of aqueous GO suspensions changes from a Newtonian viscous fluid to a shear-thinning, yield stress fluid with increasing nanoparticle concentration\cite{Corker2019}. Due to the high aspect ratio of the nanosheets the suspension gelates at much lower concentrations compared to spherical particles \cite{barwich2022}. Furthermore, our model fluid has the advantage that the surface tension is independent of the particle concentration across the whole range of rheological behaviour which facilitates the identification of the viscous and surface tension contributions to the spreading ratio $\beta_{max}$.  

We investigate the spreading dynamics and spreading ratio $\beta_{max}$ at varying concentrations and impact velocities for GO in ethanol-water suspensions using high speed photography.  We characterize the fluid properties of samples with varying concentration and compare the experimentally maximum spreading diameter to a model proposed for Newtonian fluids by Lee \textit{et al.}\cite{Lee2015} using a characteristic viscosity for each sample. We then develop an energy balance model that accounts for both viscous dissipation and the change in surface energies during spreading process. While the spreading behaviour is dominated by viscous dissipation at all concentrations, we find that there is a non-negligible contribution from the surface energies. Our model prediction for the spreading ratio $\beta_{max}$ is in good agreement with our data and compares well with literature data for Newtonian fluids.    

\section{\label{sec:methods}Experimental methods}

\begin{figure}
  \centering
  \includegraphics[height=5cm]{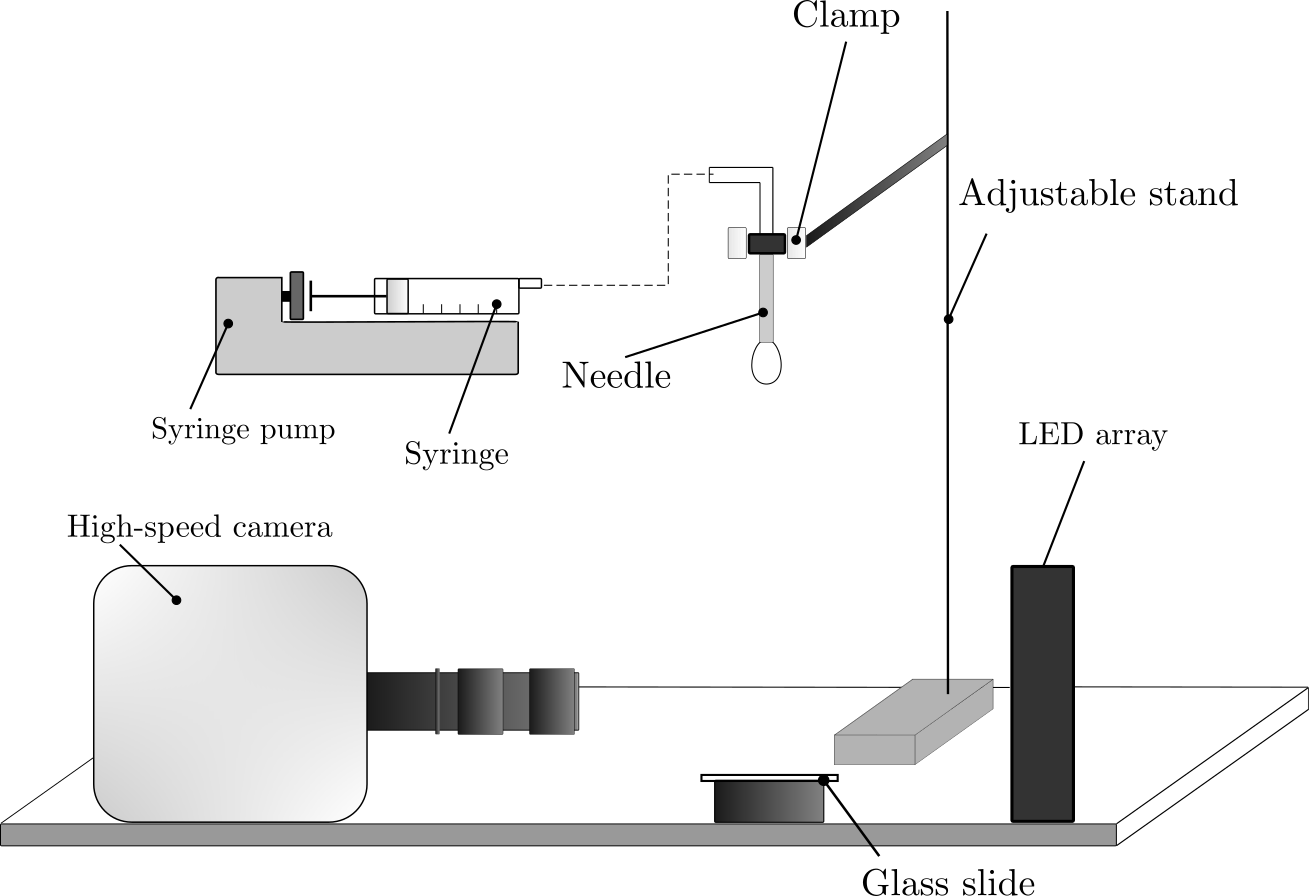} 
  \caption{Experimental apparatus for imaging droplet impacts. }
  \label{fgr:apparatus}
\end{figure}
\subsection{Sample preparation and characterisation}

\begin{table*}[ht]
\small
  \caption{\ Fluid parameters, model parameter $\alpha$ in energy balance model [Eq. (\ref{Eqn:EBM})] and Herschel-Bulkley parameters for samples of GO in 50\%v/v ethanol-water (fits in Fig. \ref{fgr:rheo}). For all GO concentrations, the density $\rho=0.92$ g/ml, the droplet radius before impact $R_0=1.3$ mm and the equilibrium contact angle $\theta_{eq}=20 \pm 5^{\circ}$.}
  \label{tbl:example1}
  \begin{ruledtabular}
  \begin{tabular*}{\textwidth}{@{\extracolsep{\fill}}lllllllll}
     GO concentration (g/l) &
    \multicolumn{1}{c}{Model parameter} &
    \multicolumn{3}{c}{Herschel-Bulkley parameters}\\
     \cmidrule(lr){3-5} 
      {} &  $\alpha$ & $\tau_0 (mPa)$ & $k (mPa\text{ }s^n)$ & $n$ \\
    \hline
    0    & 0.8 &0      &2.9  &1\\
    0.5  & 1.0 &0      &3.3  &0.987\\
    1    & 1.2 &0      &3.9  &0.965\\
    1.5  & 1.0 &0.2    &4.8  &0.950\\
    2    & 0.8 &0.5    &6.1  &0.955\\
    3    & 1.0 &3.0    &12.7 &0.906\\
    4    & 1.2 &12.5   &32.9 &0.822\\
    5    & 1.0 &70.0   &72.8 &0.774\\
    6    & 0.8 &290    &137  &0.758\\
  \end{tabular*}
  \end{ruledtabular}
  \label{table:CAHB}
\end{table*}

\begin{figure}[h]
  \centering
  \includegraphics[width=7cm]{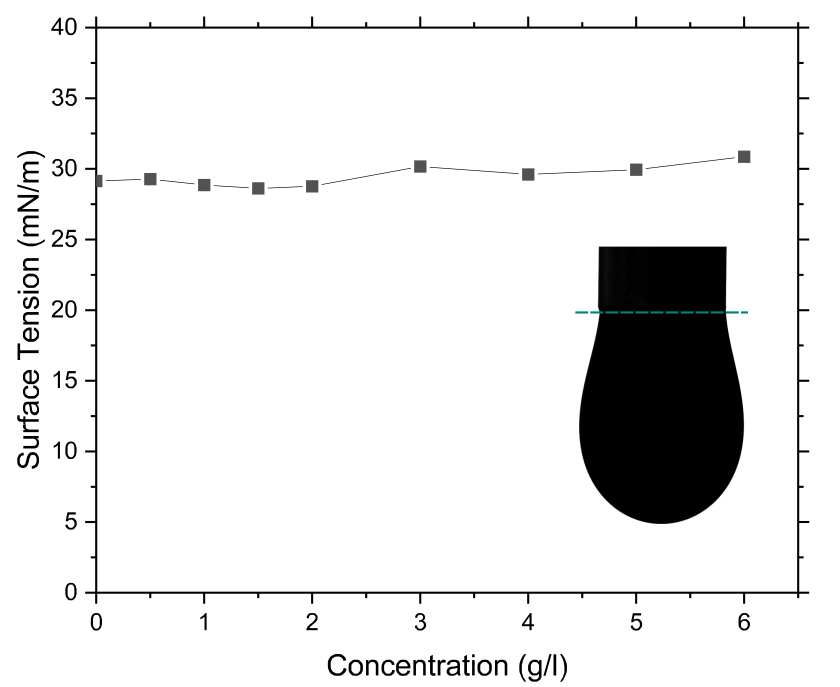} 
  \caption{Surface tension as a function of GO concentration in 50\%v/v water-ethanol suspensions. Inset shows pendant drop of g/l GO in 50\%v/v ethanol-water. The dotted line denotes the boundary between the drop and nozzle. }
  \label{fgr:ST}
\end{figure}

\begin{figure}[h]
  \centering
  \includegraphics[width=7.5cm]{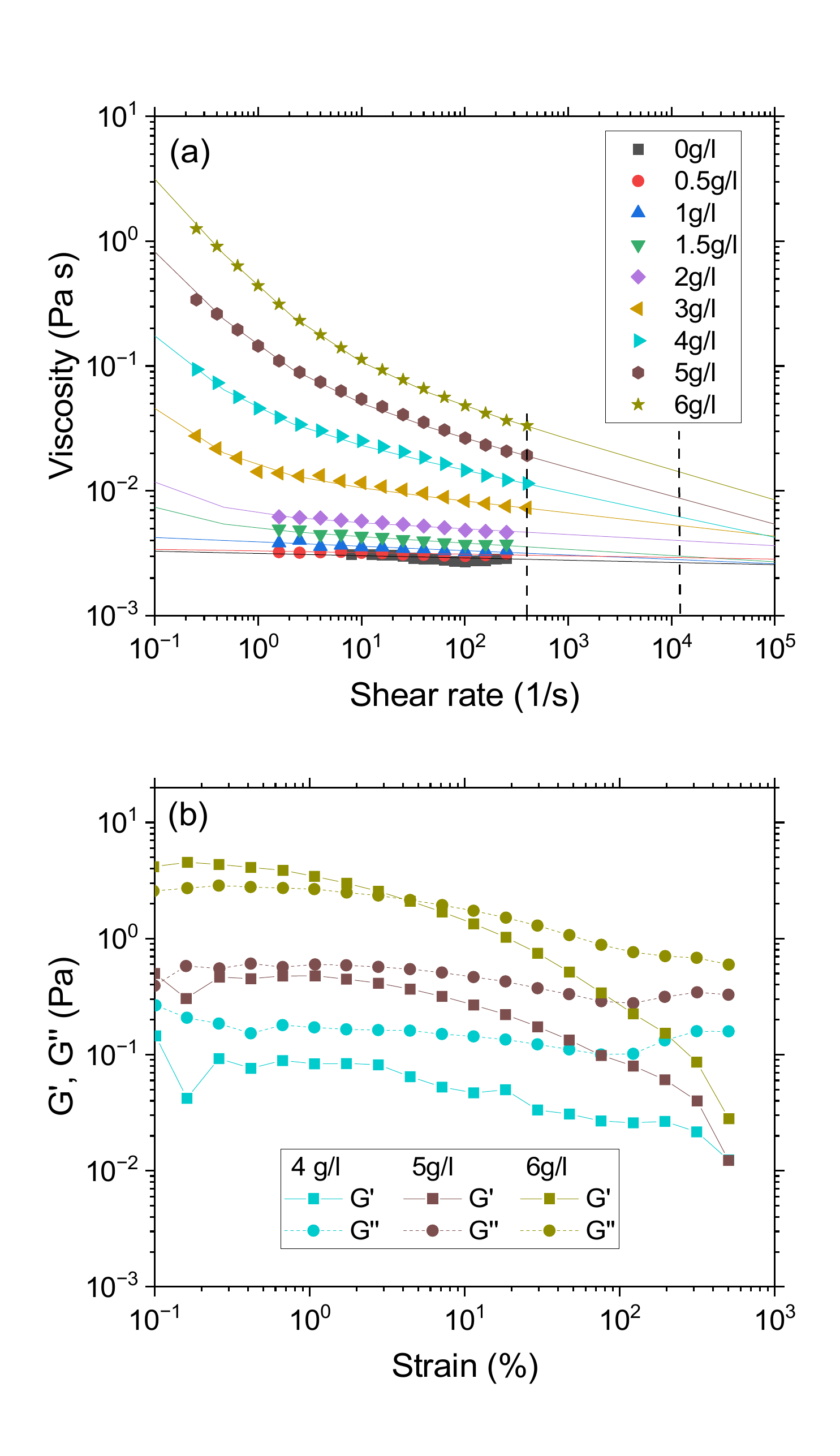} 
  \caption{(a) Viscosity as a function of shear rate for varying concentrations of GO in 50\%v/v ethanol-water. Herschel-Bulkley fits to the data is indicated by the solid lines which are extended to estimated maximum shear rates present in the droplet impact experiments. The vertical dashed lines represents the minimum and maximum shear rates reached during the droplet impact experiments, estimated using Eqs. (\ref{Eqn:shear rate}) and (\ref{Eqn:h}). (b) Strain sweep at frequency 1Hz for the three highest concentrations.}
  \label{fgr:rheo}
\end{figure}

GO samples were prepared by the dilution of a highly concentrated 2.5wt\% stock of GO in water, commercially supplied by Graphenea, USA. The stock has a mean lateral flake size of $15\pm5$ $\mu$m, verified by SEM imaging \cite{barwich2022}. The thickness of the sheets is $0.8$ nm, giving rise to an aspect ratio of $18750$. Samples were diluted with ethanol and water to produce dispersions in a 50\%v/v ethanol-water mixture. The pH of each sample was adjusted to $7$ by the addition of $0.5$M NaOH to enhance the stability of the suspension\cite{wu2013}. After dilution, all samples were shear mixed at $6000$rpm for $30$ mins in a Silverson L5M-A shear mixer to ensure the GO is well dispersed. Samples were prepared in a range of concentrations from $0-6$ g/l GO in 50\%v/v ethanol water, which remained stable for several days.

Surface tension measurements were carried out using a Dataphysics OCA 25 tensiometer using the pendant drop method on droplets suspended from $2.08$ mm outer diameter nozzles. The accompanying Dataphysics software was used to analyse the drop shape to determine the surface tension. The OCA 25 was also used to measure the equilibrium contact angle from sessile droplets.
Rheological measurements were carried out using an Anton Paar MCR 302e with a $50$ mm cross-hatched plate-plate geometry at a gap size of $1$ mm enclosed in a solvent trap. Shear controlled flow ramps were carried out with shear rates from $0.2$ to $400$  $\text{s}^{-1}$, measuring the apparent viscosity. In addition, we performed strain sweeps for the higher concentration samples. 

\subsection{Droplet impact experiments}

Drops were ejected from a $0.8$ mm outer diameter blunt tip stainless steel needle using a syringe pump, with the experimental set up shown in Fig.\ref{fgr:apparatus}. The drops impacted on a clean glass slide, with the fall height adjusted to vary the impact velocity of the impinging drop with radius $R_0=1.3$mm. Impact velocities range from $0.3-3$ m/s with impacts below the splashing transition only considered. Impacts were backlit with a Phylox HSC LED array and captured using a Phantom v2010 highspeed camera at a frame rate of $40,000$ frames per second. A custom MATLAB script was created to measure the droplet diameter and impact velocity of each drop. Spreading diameters were tracked up to $30$ ms after the droplet impact for non-splashing droplets using a custom MATLAB routine. The dynamic contact angle during spreading was measured manually with ImageJ.

An additional droplet impact measurement was carried out for the $0$g/l sample using a MD-P-801 autodrop system that dispenses droplets by means of a piezo-driven $70\mu$m pipette to study the spreading of a much smaller droplet with radius $R_0=45\mu$m and impact velocity of $1.7$ m/s.

\section{Results}
\subsection{Surface tension and rheology}
Pendant drop measurements for samples of varying GO concentration in 50\%v/v ethanol-water indicated that the surface tension of such suspensions does not vary with an increase in particle loading. Each data point in Fig. \ref{fgr:ST} represents an average of $5$ independent measurements for each concentration of GO. The surface tension across all concentrations is $30\pm1$ mN/m.

Fig. \ref{fgr:rheo}(a) shows the viscosity data of GO in 50\%v/v ethanol-water for a range of shear rates up to $400\text{ s}^{-1}$. The shear thinning behaviour of the suspensions can be clearly observed from the plot, with the low shear viscosity increasing with increased GO concentration. Below $\approx 1.0$ g/l the suspension is Newtonian with no appreciable shear rate dependence and above it exhibits shear thinning behaviour, which is fit to a Herschel-Bulkley model.  
\begin{equation}
  \eta = \frac{\tau_{HB}}{\dot{\gamma}};\quad \text{where} \quad \tau_{HB} = \tau_0 + m {\dot{\gamma}}^n \label{eqn:HB}
\end{equation}
where $\tau_0$, $m$, $n$ and $\dot{\gamma}$ are the yield stress, consistency, flow index and shear rate respectively. These fits are indicated in Fig. \ref{fgr:rheo}(a) by the solid lines. The fit paramaters are listed in table \ref{table:CAHB}

The gelation point of these suspensions which occurs at a GO concentration of around 5g/l which can be seen from the strain sweep in \ref{fgr:rheo}(b). The storage modulus G' becomes larger than G'' for concentrations larger than 5g/l at low strain amplitudes. From the Herschel Bulkley fits, a non-zero yield stress emerges at concentrations above 1.5 g/l.

\begin{figure}[h]
\centering
  \includegraphics[height=6.5cm]{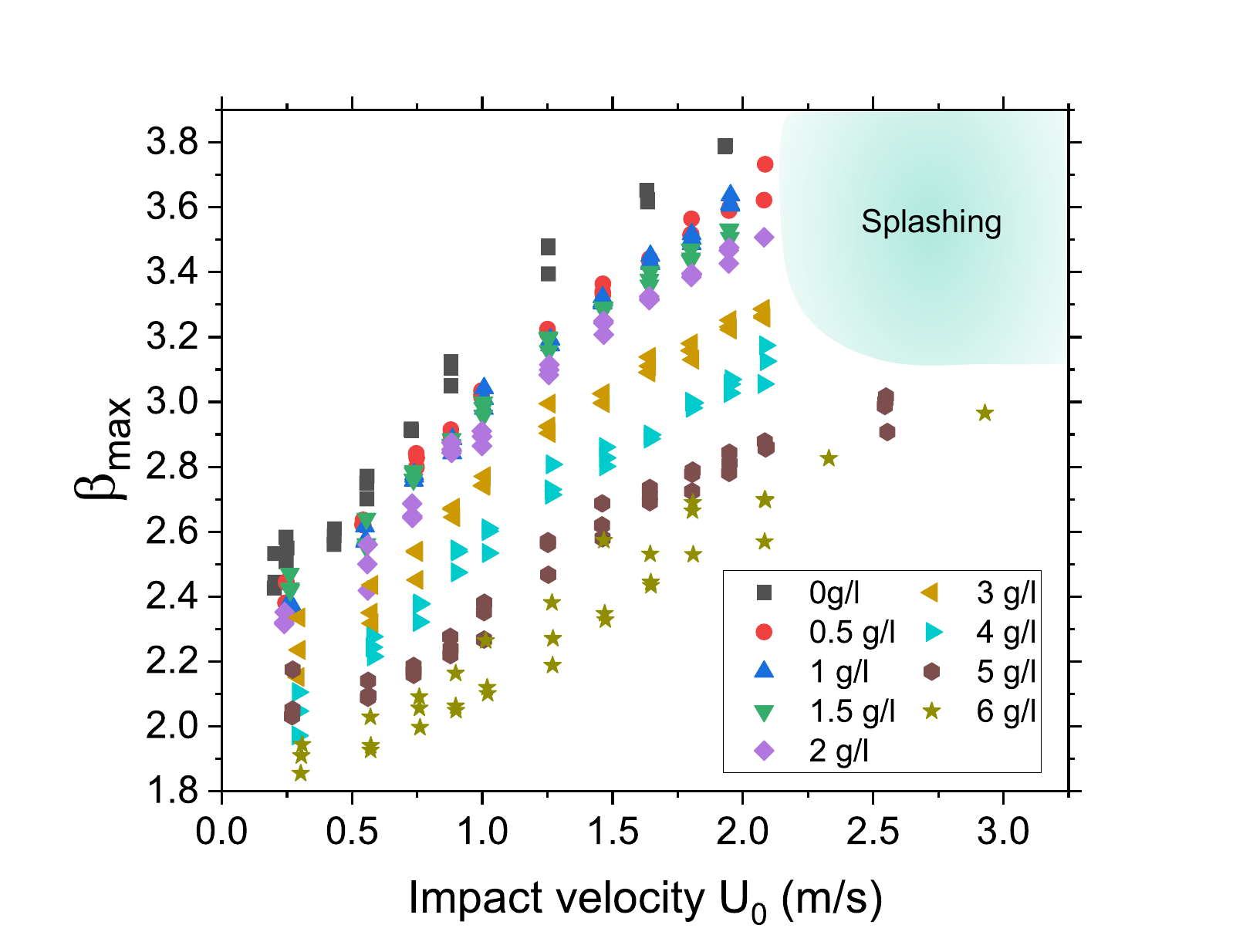} 
  \caption{Spreading ratio $\beta_{max}$ as a function of impact velocity $U_0$ for varying GO concentration in 50\%v/v ethanol water.}
  \label{fgr:Bmax vel}
\end{figure}

\begin{figure*}[!htbp]
  \includegraphics[width=0.8\textwidth]{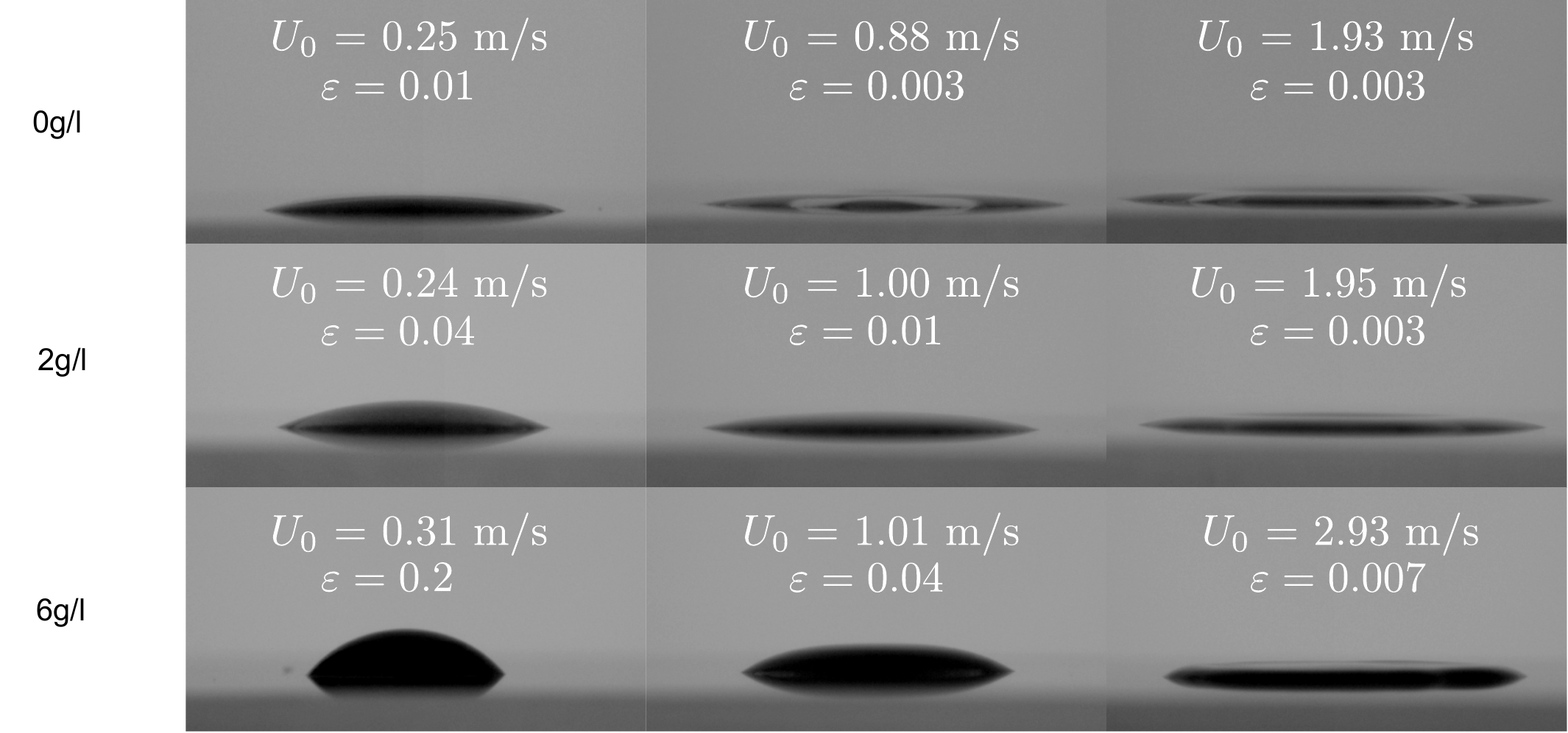}
  \caption{Droplet shape at $\beta_{max}$ and corresponding shape parameter $\epsilon$ [Eq.(\ref{Eqn:epsilon})] for the lowest, an intermediate and the highest impact velocity $U_0$ as indicated at 3 different concentrations: $0$g/l, $2$g/l, $6$g/l. The spherical cap approximation was used to calculate $\epsilon$.}
  \label{fgr:image plate}
\end{figure*}

\subsection{Spreading ratio}

When a drop impacts on a solid substrate the drop flattens and spreads tangentially across the substrate until it reaches some maximum spreading diameter. For a sufficiently high impact velocity, drops can be ejected from the spreading drop resulting in splashing phenomena. The tangential spreading is inertia driven with capillary and viscous forces opposing the motion and non-dimensional numbers are typically used to describe the balance between these forces. The Weber number, $We = \rho (2R_0) {U_0}^2/\sigma$, describes the ratio of inertial to capillary forces and the Reynolds number, $Re = \rho (2R_0) U_0/\mu$, describes the ratio of inertial to viscous forces, where $\rho$, $R_0$, $U_0$, $\sigma$, and $\eta$ are the density, initial droplet radius, impact velocity, surface tension and viscosity respectively.

In order to account for the shear-thinning properties of the GO suspensions, we adapt the method used by Balzan \textit{et al.}\cite{Balzan2021a} to calculate a characteristic viscosity during the spreading process. Using an approximation for the shear rate during spreading \cite{Chandra1991}

\begin{equation}
    \dot{\gamma} \sim \frac{U_0}{h} \label{Eqn:shear rate},
\end{equation}  where $h$ the lamella thickness at maximum spread which can be approximated as a cylindrical disc with radius $R$. From volume conservation we obtain
\begin{equation}
    h = \frac{4 {R_0}^3}{3 R^2} \label{Eqn:h}
\end{equation}
where $U_0$ and $h$ are the impact velocity and lamella thickness at maximum spread respectively.
We deduce an expression for a characteristic viscosity based on a Herschel-Bulkley model using Eqns (\ref{eqn:HB}), (\ref{Eqn:shear rate}) and (\ref{Eqn:h}),

\begin{equation}
    \eta_{HB} = \tau_0 \left(\frac{4 {R_0}^3}{3 U_0 R^2}\right) + k\left(\frac{3 U_0 R^2}{4 {R_0}^3}\right)^{n-1}. \label{eqn:muhb}
\end{equation}

We use this to define a generalized Reynolds number, 
\begin{equation}
Re=\frac{\rho (2R_0) U_0}{\eta_{HB}}
\label{Eqn:GRe}
\end{equation}
which depends not only on $U_0, R_0$ and $\rho$ but also on $R$ and the Herschel-Bulkley fit parameters [Fig. \ref{fgr:rheo}(a), table \ref{table:CAHB}] via Eq. (\ref{eqn:muhb}). It reduces to the ordinary Reynolds number for Newtonian fluids.

Experiments observing the maximum spreading diameter were carried out up to velocities below the splashing threshold for samples of varying GO concentration. The spreading ratio, $\beta_{max} = R/R_0$, of these measurements as a function of impact velocity $U_0$ is shown in Fig. \ref{fgr:Bmax vel}. As the GO concentration is increased there is a clear decrease in the spreading ratio across all impact velocities particularly for samples above $2$ g/l. 

Fig. \ref{fgr:image plate} shows the droplet shape at maximum spread at different impact velocities $U_0$ for three different GO concentrations in 50\%v/v ethanol-water. The strong influence of impact velocity on both spreading radius, droplet shape and contact angle is can be clearly observed. At low concentrations and high impact velocities the shape is disc-like, while at lower impact velocities and higher concentrations it resembles a spherical cap.

There are numerous predictions for the maximum spreading ratio of Newtonian fluids based on semi-empirical \cite{Lee2015,Laan2014, Scheller1995,Madejski1975, Bennett1993, Asai1993, Mao1997, Roisman2009}, numerical \cite{Wildeman2016,Healy2001} and energy balance models \cite{aksoy2022,Chandra1991,Pasandideh-Fard1996,Mao1997,Ukiwe2005,Wildeman2016,huang2018,du2021,worner2023, Zhao2018}. 
Studies on Newtonian liquids to predict $\beta_{max}$ typically distinguish between the viscous and capillary regimes. Clanet \textit{et al.}\cite{Clanet2004} have shown that when viscous dissipation dominates the spreading ratio scales as $\beta_{max} \sim {Re}^{1/5}$ which is supported by several experimental studies\cite{Roisman2009, Madejski1975, Clanet2004,Fedorchenko2005}.
In the capillary regime, the spreading ratio scales as $\beta_{max} \sim {We}^{1/2}$ which is derived from energy conservation between kinetic energy and surface energy \cite{Bennett1993, Eggers2010, Chandra1991}.

An impact parameter $P$ proposed by Eggers \textit{et al.}\cite{Eggers2010} may be used to determine in which regime – capillary or viscous - our data falls.

\begin{figure}[ht]
\centering
  \includegraphics[height=7cm]{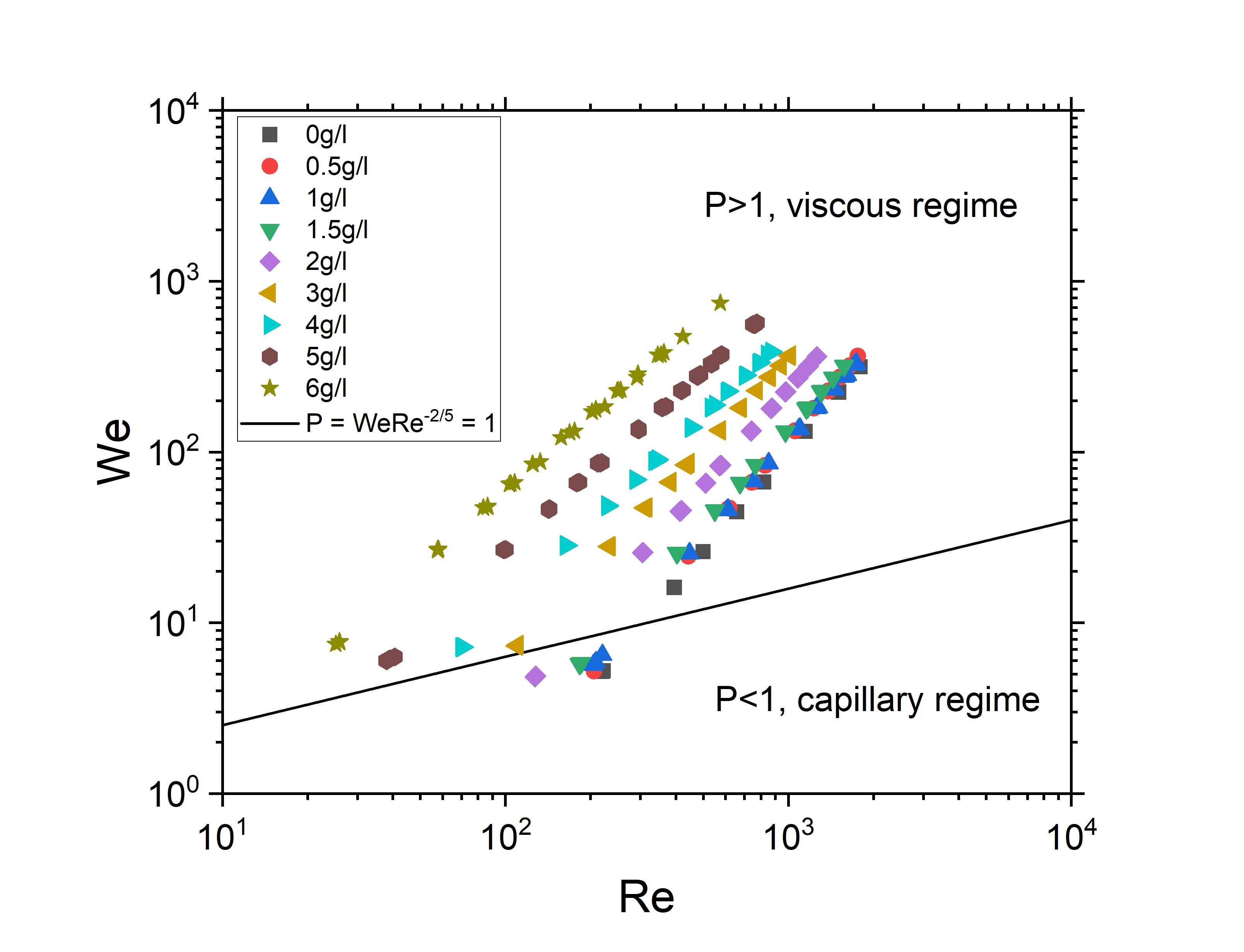}
  \caption{Phase diagram of impact regimes defined by impact parameter\cite{Eggers2010, Laan2014} $P=We{Re}^{-2/5}$.}
  \label{fgr:Regime}
\end{figure}

It is defined as
\begin{equation}
    P \equiv We{Re}^{-2/5}
\end{equation}
If $P\gg1$, viscous dissipation dominates, while for $P\ll1$ it is the change in surface energy. Almost all experimental data for our GO suspensions including most of the 0g/l data fall in the viscous regime, as shown in Fig. \ref{fgr:Regime}. Only for low concentration and low impact velocity surface forces are relevant.

\begin{figure}[h]
\centering
  \includegraphics[height=6.5cm]{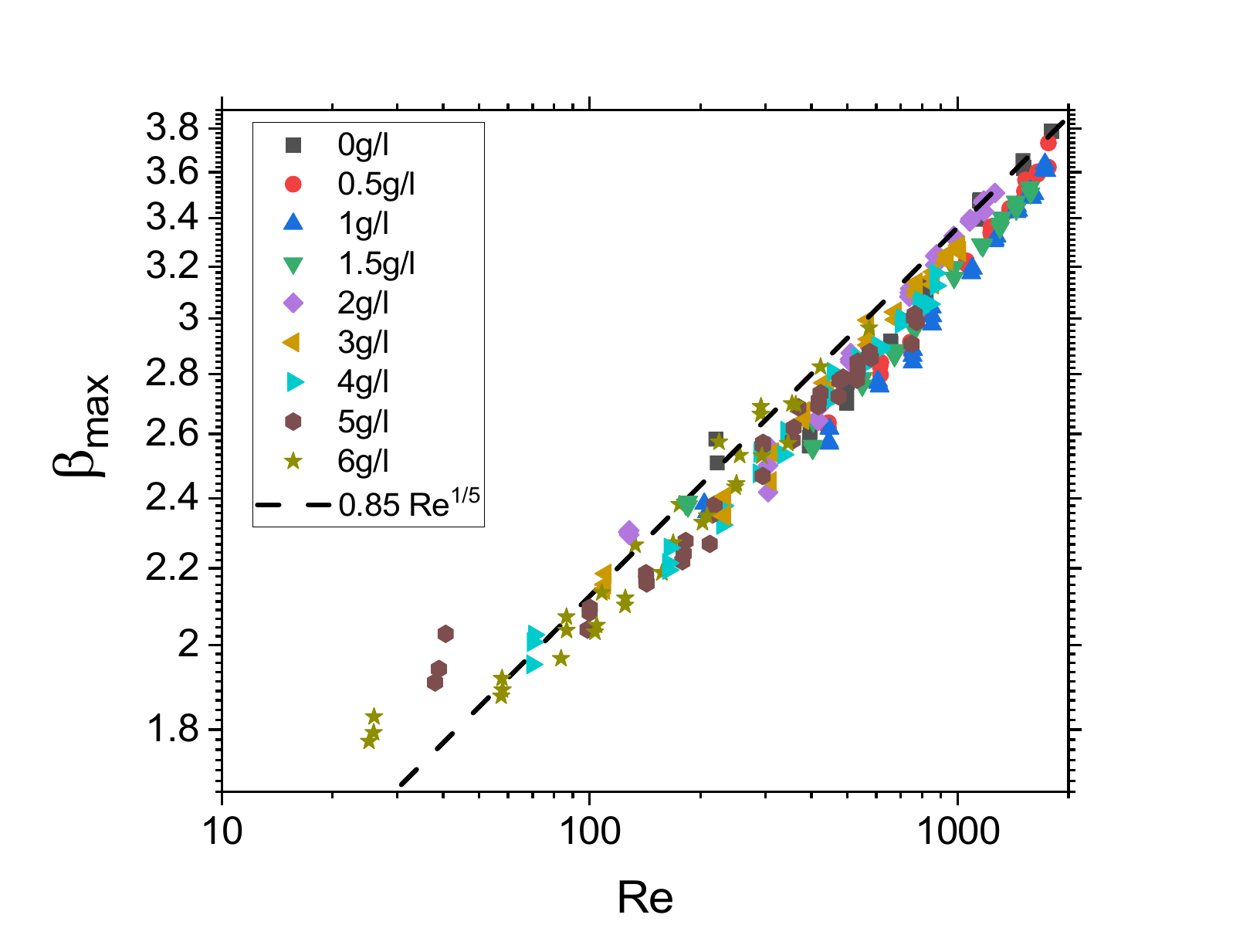}  
  \caption{$\beta_{max}$ dependence on $Re$ for all concentrations of GO in 50\%v/v ethanol-water. Dotted line represents the high viscosity, low surface tension limit of the our energy balance model, $\beta_{max} = 0.85 Re^{1/5}$ [Eq. (\ref{Eqn:visclimit})].}
  \label{fgr:Bmax Re}
\end{figure}

This can also be seen by plotting $\beta_{max}$ versus the generalised Reynolds number $Re$ [Eq. (\ref{Eqn:GRe})] as shown in Fig. \ref{fgr:Bmax Re}. At all concentrations the data follows to a good approximation the viscous $Re^{1/5}$ scaling though deviations are observed for lower $Re$, where surface tension forces cannot be neglected.  

\begin{figure}[h]
\centering
  \includegraphics[height=6.5cm]{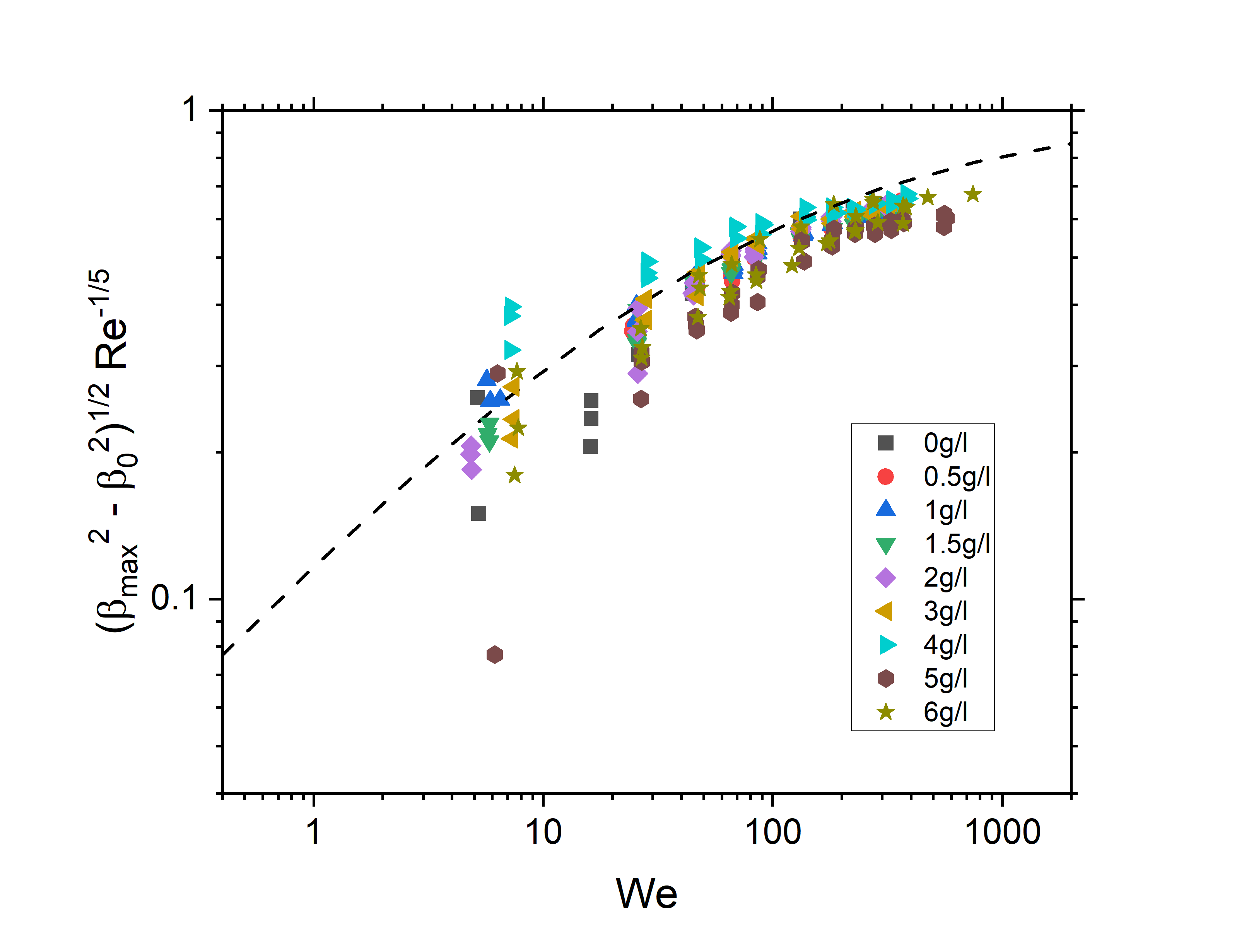}
  \caption{Comparison of GO experimental data to the spreading model of Lee \textit{et al.} [Eq.(\ref{Eqn:Lee})] with $A=7.6$ (dashed line) where the values for $\beta_0$ for each concentration were found using the fit equation Eq. (\ref{Eqn:beta0}).}
  \label{fgr:Lee model}
\end{figure}

Lee \textit{et al.} proposed a model that interpolates between the viscous and capillary regimes using a Pad\'{e} approximation. In order to account for the non-zero spreading ratio in the limit of zero impact velocity \cite{Lee2015}, they introduce the limiting value $\beta_0$ obtained by fitting the experimental data to the function
\begin{equation}
    \beta_{max} = \beta_0 + A \frac{{U_0}^C}{B + {U_0}^C}, \label{Eqn:beta0}
\end{equation}
where $\beta_0, A, B, C$ are all fitting parameters. Using $\beta_0$ and the Pad\'{e} approximation they arrive at a relation between the spreading ratio and the two scaling regimes that captures the spreading behaviour of a wide range of Newtonian fluids with a fit parameter $A=7.6$.
\begin{equation}
    \left({\beta_{max}}^2 - {\beta_{0}}^2 \right)^{1/2} {Re}^{-1/5} = \frac{{We}^{1/2} }{A + {We}^{1/2} }.
    \label{Eqn:Lee}
\end{equation}
Comparison of our experimental data using the generalized Reynolds number [Eq.(\ref{Eqn:GRe})] to this model in Fig. \ref{fgr:Lee model} shows some agreement but with considerable scatter.

\subsection{Energy balance of spreading process}
Several models for Newtonian fluid droplets spreading on a smooth surface based on an energy balance have been proposed in the literature\cite{Chandra1991,Pasandideh-Fard1996,Mao1997,Ukiwe2005,Wildeman2016,huang2018,du2021,worner2023, Zhao2018}. In general, the kinetic energy of the droplet  on impact is balanced by the change in surface energy and viscous dissipation during the spreading.
\begin{equation}
    E_{kin} = \Delta E_{s} + E_v, \label{Eqn:EB1}
\end{equation}
where $E_{kin}$, $ \Delta E_{s}$ and $E_v$ are the kinetic energy, the change in surface energy and viscous dissipation. The kinetic energy of a droplet with radius $R_0$ and corresponding volume $V_d=4\pi R_0^3/3$ is simply 
\begin{equation}
  E_{kin} = \frac{1}{2}\rho V_d U_0^2=\frac{2}{3} \rho \pi R_0^3 {U_0}^2 \label{Eqn:EB2}
\end{equation}
\begin{figure}[h]
\centering
  \includegraphics[height=4cm]{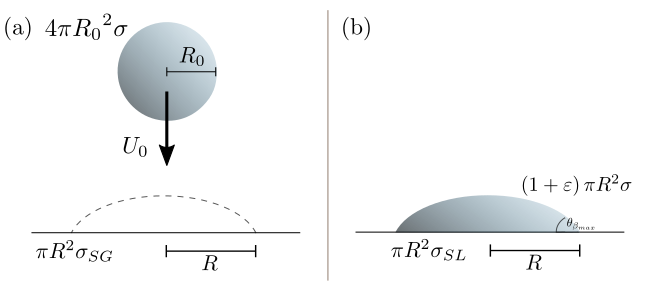}
  \caption{Surface energy contributions (a) before and (b) after impact of a spherical drop at maximum spread.}
  \label{fgr:EB diag}
\end{figure}
As shown in Fig. \ref{fgr:EB diag} the surface energy has two contributions\cite{Ukiwe2005,Ford1967}, firstly the energies of the surfaces before impact, namely the surface energy of the spherical drop and the energy between the substrate and the gas. The second contribution is the energies after spreading, the energy of the interface between the drop and the surrounding gas at maximum spread and the energy between the substrate and the drop. A priori, the shape and therefore the interfacial area $A_{\beta_{max}}$ between droplet and gas at maximum spread is unknown and different approximations have been made in the literature. In general we can express this area as
\begin{equation}
  A_{\beta_{max}}=(1+\epsilon)\pi R^2
  \label{Eqn:epsilon}
\end{equation}
where $\epsilon$ is a shape parameter. At high impact speeds and large $\beta_{max}$, the droplet is often pancake shaped and the corresponding area well approximated by either a simple disc\cite{Chandra1991,Wildeman2016} with $\epsilon=0$ or a thin cylinder \cite{du2021,Pasandideh-Fard1996,Ukiwe2005, aksoy2022}, where $\epsilon=8/3\beta_{max}^3$, which can be derived from mass conservation [Eq. (\ref{Eqn:h})]. At lower impact speeds, the shape can be approximated by a spherical cap 
\begin{equation}
  \epsilon=4\sin^2(\theta_{\beta_{max}}/2)/\sin^2(\theta_{\beta_{max}})-1,
  \label{Eqn:cap}
\end{equation} 
where $\theta_{\beta_{max}}$ is the contact angle at maximum spread. However deviations from that shape have been observed \cite{Lee2015}. Note that for contact angles $\theta_{\beta_{max}}\ge 90^{\circ}$, the shape parameter $\epsilon \ge 1$.  Here we keep $\epsilon$ as an adjustable constant. This may appear unjustified as $\epsilon$ is a function of $\beta_{max}$ and decreases towards zero for large $\beta_{max}$ where the droplet shape resembles a thin disk. However, at large $\beta_{max}$ the viscous contribution dominates the energy balance as the viscous dissipation scales as $E_v \propto \beta_{max}^5$ [Eq. (\ref{Eqn:EB4})] compared to the surface energy contributions $\Delta E_s\propto \beta_{max}^2$ [Eq. (\ref{Eqn:EB3b})]. In this regime, the value of $\epsilon$ plays no role. Therefore, $\epsilon$ reflects the droplet shape at low impact velocities where it approaches a constant.

The change in surface energy is then
\begin{eqnarray}
     \Delta E_s &=& E_{s,final}-E_{s,inital} \\
     &=& \left( \pi R^2 \sigma_{SL} + (1+\epsilon) \pi R^2 \sigma  \right)-\left(  4 \pi R_0^2 \sigma + \pi R^2 \sigma_{SG} \right),  \label{Eqn:EB3a}
\end{eqnarray}
where R is the radius of the drop at maximum spread, $\sigma_{SG}$ is the interfacial tension between the substrate and the gas and $\sigma_{SL}$ is the interfacial tension between the substrate and the drop.

Using the definition of the spreading coefficient 
\begin{equation}
   S = \sigma_{SG} - \sigma_{SL} - \sigma
     = \sigma \left(  \cos{\theta} -1  \right),
\end{equation}
Eq. (\ref{Eqn:EB3a}) can be simplified to
\begin{equation}
    \Delta E_s = \pi R_0^2 \sigma \left( \alpha \beta_{max}^2 -4  \right) .
    \label{Eqn:EB3b}
\end{equation}
where 
\begin{equation}
  \alpha=\epsilon+(1-\cos{\theta_d})
  \label{Eqn:alpha}
\end{equation}

The parameter $\alpha$ combines the shape parameter $\epsilon$ with the contact angle $\theta_d$. It is important to note that in the context of spreading, $\theta_d$ does not correspond to the equilibrium contact angle $\theta_{eq}$ but rather the average dynamic contact angle during spreading $\theta_d$ 
as it reflects the work done moving the contact line\cite{sikalo2005,vadillo2009}. This dynamic contact angle can be significantly larger\cite{Lee2015} than $\theta_{eq}$ and also depends on the fluid parameters\cite{vadillo2009}.

The last term is the viscous dissipation energy, given by\cite{Chandra1991}
\begin{equation}
    E_v =  \eta_{HB} {\dot{\gamma}}^2 \tau_s V_d \label{Eqn:Ev}
\end{equation}
for drop volume $V_d$ and time to maximum spread $\tau_s$, which we can estimate to be 
\begin{equation}
  \tau_s \sim \frac{R}{U_0} \label{Eqn:taus}
\end{equation}
\begin{figure*}[!htbp]
  \includegraphics[width=\textwidth]{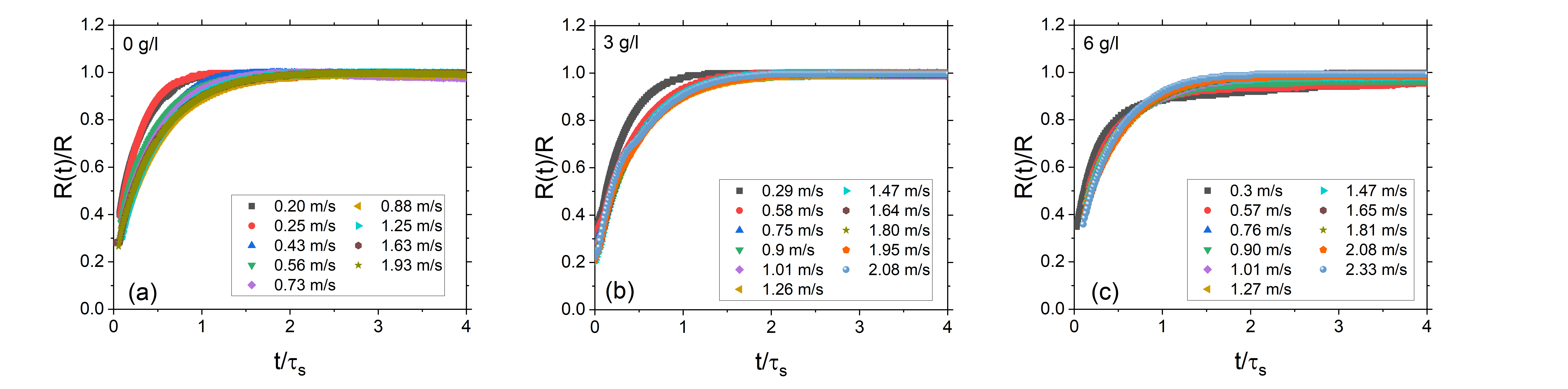} 
  \caption{Time evolution of normalised spreading radius $R(t)/R$ versus time rescaled by the spreading time $\tau_s$ [Eq.(\ref{Eqn:taus})] at different impact velocities for 3 different concentrations: (a) 0g/l, (b) 2g/l and (c) 6g/l.}
  \label{fgr:spreadingtime}
\end{figure*}
In order to assess the validity of $\tau_s$, we measured the time evolution of the normalized spreading radius $R(t)/R$ versus the time rescaled by $\tau_s$ as shown in Fig. \ref{fgr:spreadingtime}. Recent experiments by Gorin \textit{et al.}\cite{gorin2022} showed that this type of scaling follows a universal curve for a wide range of Newtonian and non-Newtonian fluids. In our case most of the data does collapse but shows deviations at low impact velocities. Nevertheless, we note that in all cases the spreading radius reaches at least $\ge$90\% of its maximum value at $t=\tau_s$. Therefore $\tau_s$ represents a fair estimate of the spreading time.     

Using the shear rate assumption from Eq.(\ref{Eqn:shear rate}) with an estimate of the lamella thickness from volume conservation from a spherical droplet to a disc [Eq.(\ref{Eqn:h})] the viscous dissipation [Eq. (\ref{Eqn:Ev})] is given by
\begin{equation}
    E_v =  \frac{3}{4}\pi \eta_{HB} U_0 R_0^2 {\beta_{max}}^5 . \label{Eqn:EB4}
\end{equation}
This estimate for viscous dissipation is similar to that of Chandra \textit{et al.}\cite{Chandra1991}. The difference is that the authors used the spreading time scale $R_0/U_0$ rather than Eq.(\ref{Eqn:taus}) which leads to $E_v\propto \beta_{max}^4$ compared to the $E_v\propto \beta_{max}^5$ scaling in our model.  
Combining Eqs. (\ref{Eqn:EB1}), (\ref{Eqn:EB2}), (\ref{Eqn:EB3b}) and (\ref{Eqn:EB4}) we arrive at an equation relating the maximum spreading ratio to $Re$ and $We$.
\begin{equation}
    \frac{3}{4}\frac{We}{Re}{\beta_{max}}^5 +  \alpha {\beta_{max}}^2 = \frac{1}{3}We + 4, \label{Eqn:EBM}
\end{equation}
Solutions to this polynomial can be obtained numerically for a given $\alpha$ bearing in mind that the generalised Reynolds number $Re$ [Eq.(\ref{Eqn:GRe})] depends on $\beta_{max}$ via Eq.(\ref{eqn:muhb}). In the limit of high viscosity and low surface tension, where $\frac{Re}{We} \rightarrow 0$, this expression reduces to 
\begin{equation}
  \beta_{max} = \left ({\frac{4}{9}Re} \right )^{\frac{1}{5}}\label{Eqn:visclimit}
\end{equation}
The prefactor of this equation is $0.85$ and close to the value of $1.11$ that has been found by Fedorchenko\textit{et al.}\cite{Fedorchenko2005}. Fig. \ref{fgr:Bmax Re} shows this relation compared to the GO data and can be seen to give a good representation of the data though it does overestimate $\beta_{max}$ at high $Re$ and underestimate it at low $Re$ where surface tension forces matter.

In the capillary limit where $\frac{Re}{We} \gg 1$, this reduces to  
\begin{equation}
  \beta_{max}=\frac{1}{\sqrt{\alpha}}\sqrt{\frac{We}{3}+4}\sim {We}^{1/2}.
  \label{Eqn:caplimit}
\end{equation} 

For Newtonian fluids, the generalized Reynolds number reduces to the ordinary one with constant viscosity $\eta$. In this case the energy balance model [Eq.(\ref{Eqn:EBM})] is a quadratic equation in $U_0$ and can be solved analytically. For a given $\beta_{max}$ and $\alpha$, the impact velocity is given by
\begin{equation}
  U_0=\frac{9}{16}\frac{\eta \beta_{max}^5}{\rho R_0}\left ( 1+\sqrt{1+\frac{64}{27}\frac{1}{Oh^2}\frac{\alpha \beta_{max}^2-4}{\beta_{max}^{10}}}\right )
  \label{Eqn:EBMSol}
\end{equation}
where $Oh=\sqrt{We}/Re=\eta/\sqrt{\sigma \rho 2 R_0}$ is the Ohnesorge number. Note that this analytic solution also holds for the cylinder approximation, where the shape parameter $\epsilon$ is replaced with $8/3\beta_{max}^3$ in the expression for $\alpha$ [Eq.(\ref{Eqn:alpha})]. 

\begin{figure}[h]
\centering
  \includegraphics[width=7cm]{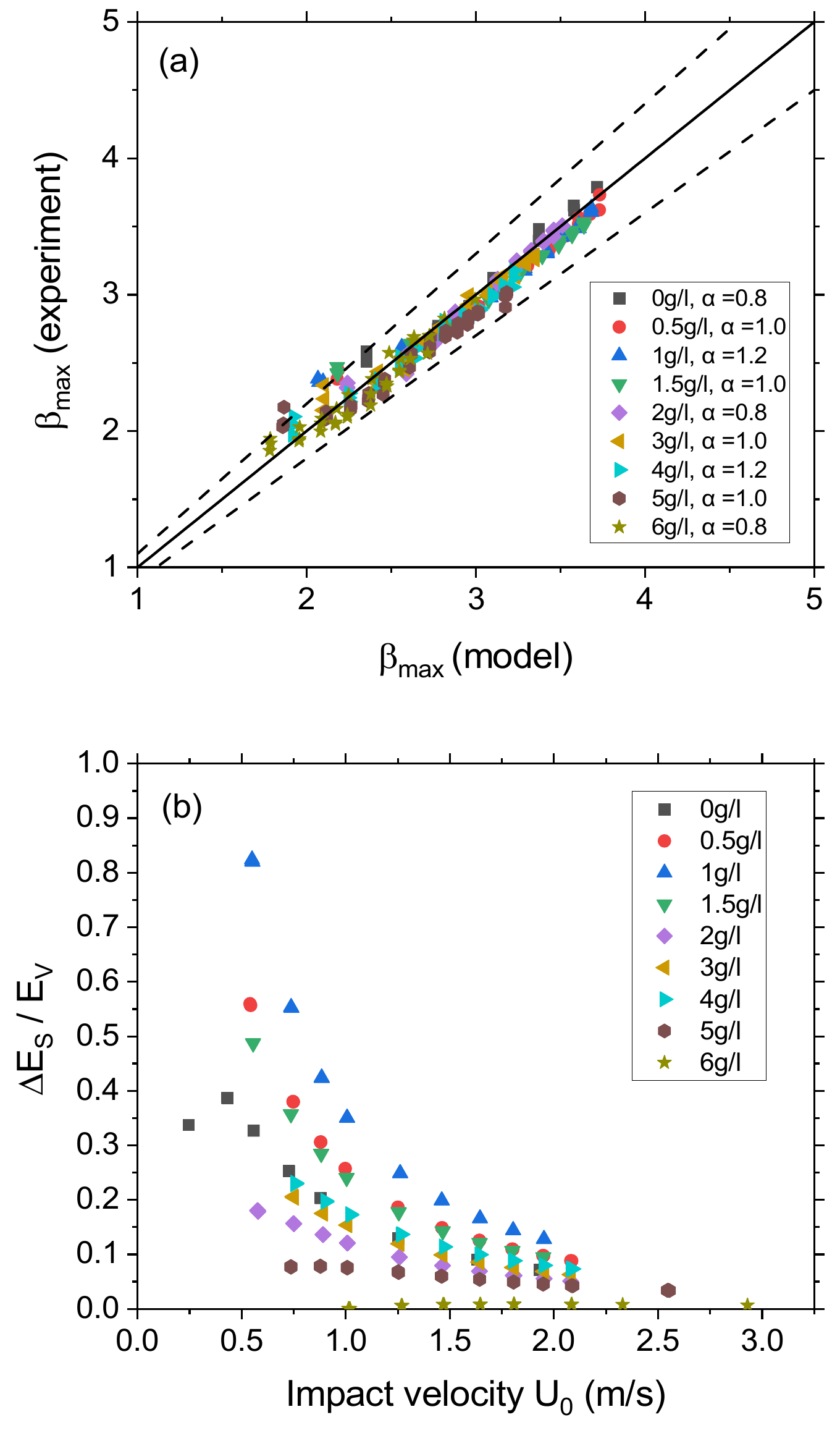}
  \caption{(a) Comparison of the GO data with our energy balance model for all concentrations with the corresponding $\alpha$ parameter that gave the best fit. Dotted lines correspond to 10\% deviations. (b) Ratio of surface energy [Eq.(\ref{Eqn:EB3b})] and viscous dissipation [Eq.(\ref{Eqn:EB4})] with $\alpha$ taken from fits in (a).}
  \label{fgr:EBM GO}
\end{figure}

\begin{figure}
\centering
  \includegraphics[height=7.0cm]{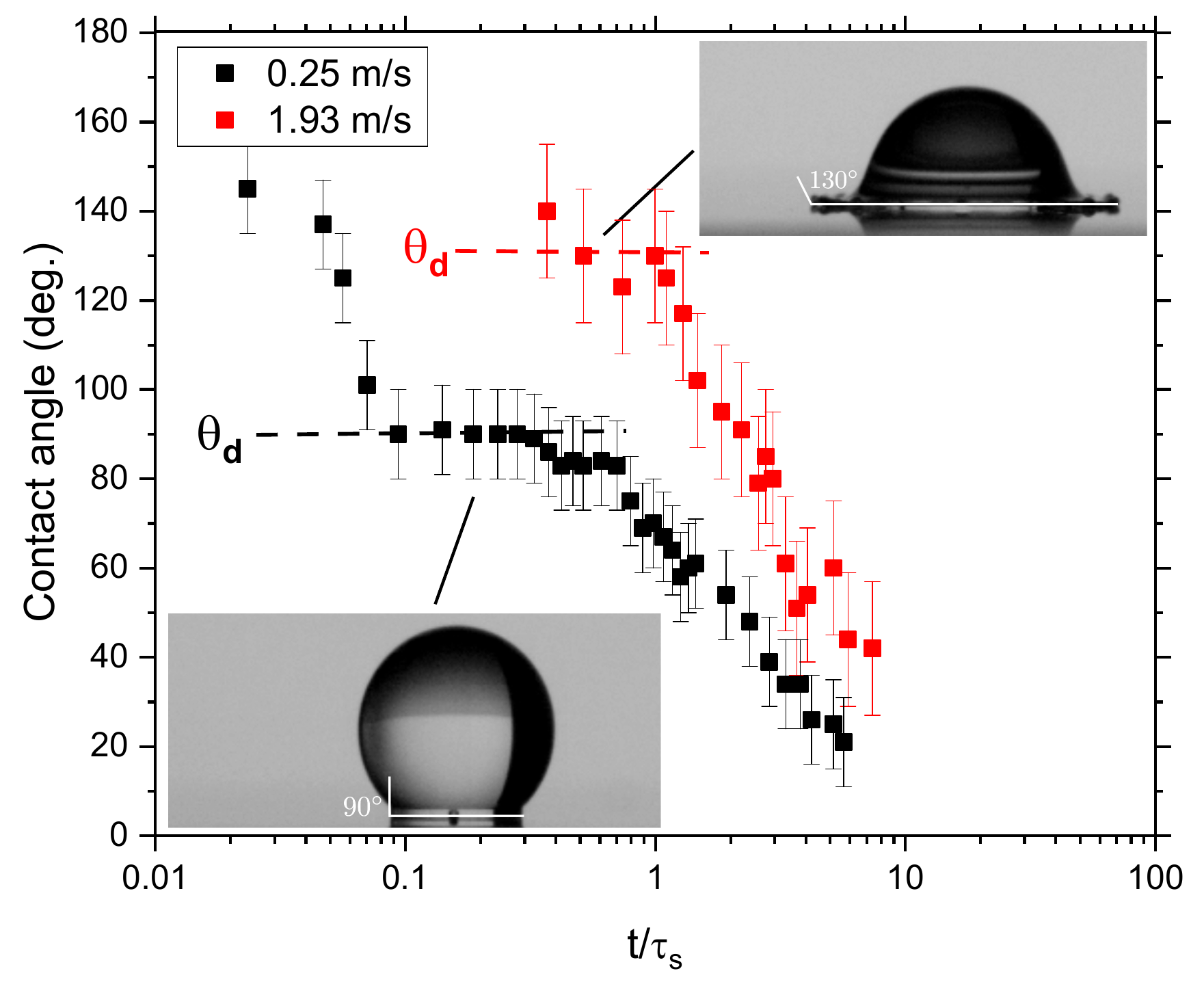}
  \caption{Dynamic contact angle as a function of time normalised by the spreading time $\tau_s$ [Eq.(\ref{Eqn:taus})] for 0g/l droplet at the lowest and highest impact velocity as indicated. The dotted line refer to the average dynamic contact angle $\theta_d$ during the spreading process.}
  \label{fgr:CA}
\end{figure}
Fig. \ref{fgr:EBM GO}(a) shows a comparison of the experimental data to model predictions from our energy balance model [Eq.(\ref{Eqn:EBM})] using $\alpha$ as a fitting parameter. These predictions are calculated using the characteristic viscosity $\eta_{HB}$ for each data point. The values for $\alpha$ do not vary much between concentrations and are between $0.8$ and $1.2$. The fits to the data are excellent and within $14\%$ of the data. 

Fig. \ref{fgr:EBM GO}(b) shows the ratio of surface energy $\Delta E_S$ [Eq.(\ref{Eqn:EB3b})] and viscous dissipation $E_v$ [Eq.(\ref{Eqn:EB4})] using the fitted values of $\alpha$ to illustrate the relative magnitude of surface tension forces compared to the viscous dissipation. As expected, the surface tension forces become less important for high impact velocities and are negligible at the highest concentration. Nevertheless, surface tension forces are appreciable for concentrations below $\sim2$g/l and impact velocities less than $\sim1$ m/s, where the ratio is $\ge 20\%$ and can reach up to $80\%$. 

In order to assess whether the value of $\alpha$ we obtained from fitting our model to the data is reasonable we measured the shape parameter $\epsilon$ and dynamic contact angle during spreading. Fig. \ref{fgr:image plate} shows the droplet shape at $\beta_{max}$ and the corresponding $\epsilon$ for different concentrations and impact velocities. Except for the highest concentration, where surface tension forces are negligible compared to viscous dissipation (see Fig. \ref{fgr:EBM GO}(b)), the shape is disk-like with $\epsilon \approx 0$. 

In Fig. \ref{fgr:CA} we track the dynamic contact angle during the spreading process for the lowest and highest impact velocity for the 0g/l fluid, where surface tension forces are non-negligible at low impact velocities (see Fig.\ref{fgr:EBM GO}(b)). Clearly, the contact angle during spreading is higher for the larger impact velocity where a thin lamella is ejected. For the low impact velocity droplet, the contact angle is not constant during spreading. It quickly drops from $145^\circ$ to around $90^\circ$ where it plateaus during the spreading process with a further drop towards the equilibrium contact angle of $20^{\circ}$ for $t>\tau_s$. Substituting the low impact velocity values, $\epsilon=0$ (Fig.\ref{fgr:image plate}) and $\theta_d\approx 90^{\circ}$ into Eq. (\ref{Eqn:alpha}) we obtain $\alpha=1$ which is consistent with the fitted $\alpha$ values that range between $0.8$ and $1.2$.

\begin{figure}
\centering
  \includegraphics[width=8cm]{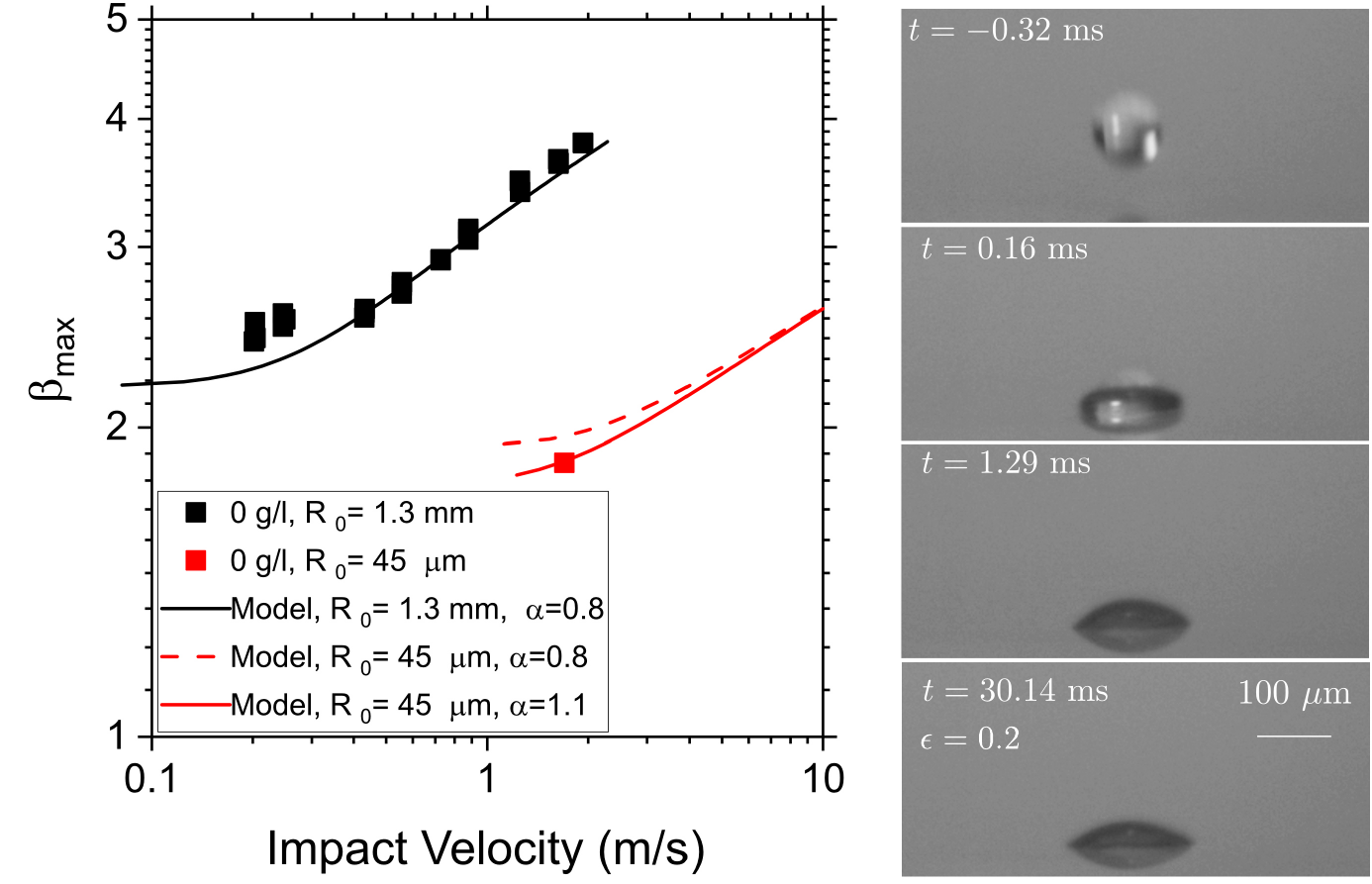}
  \caption{Spreading ratio versus impact velocity for the 0g/l sample for two different droplet sizes as indicated. Lines are the Newtonian solutions of our energy balance model [Eq. (\ref{Eqn:EBMSol})]. Image panels on the right show the snapshots of the drop impact for the $R_0=45 \mu m$ droplet with the times shown where $t=0$ is the time of impact.}
  \label{fgr:microdrop}
\end{figure}

In order to further test our model, we measured the droplet impact of a much smaller $0$g/l droplet of radius $R_0=45\mu m$ which is a factor of $\sim 30$ smaller than the droplets dispensed from the syringe. Unfortunately we could not change the impact velocity due to the limitations of the piezo-driven autodrop dispensing system we used so we only have one data point at $U_0=1.7$m/s. Snapshots of the impact at different times are shown in Fig. \ref{fgr:microdrop}. As demonstrated in Fig. \ref{fgr:microdrop} the spreading ratio for the smaller droplet  is drastically lower compared ($\beta_{max}=1.85$) to the larger droplet at the same impact velocity ($\beta_{max}=3.7$). As the wetting properties are the same in both cases we expect the value of $\alpha$ to be similar. Using the same value for $\alpha$ as found from the fit to the larger droplet data, the model prediction for the small droplet using Eq. (\ref{Eqn:EBMSol}) is within $10\%$ of the experimental result. Increasing $\alpha$ from $0.8$ to $1.1$ provides a better agreement, which is consistent with the larger value of $\epsilon$ for the small droplet which increases $\alpha$. Here, $\epsilon=0.2$ for the small droplet while $\epsilon \approx 0$ for the large droplet [Fig.\ref{fgr:image plate}]. Even though there is not enough data for a reliable fit, the model clearly captures the strong droplet size dependence of the spreading ratio. 

\begin{figure*}
  \centering
  \includegraphics[width=\textwidth]{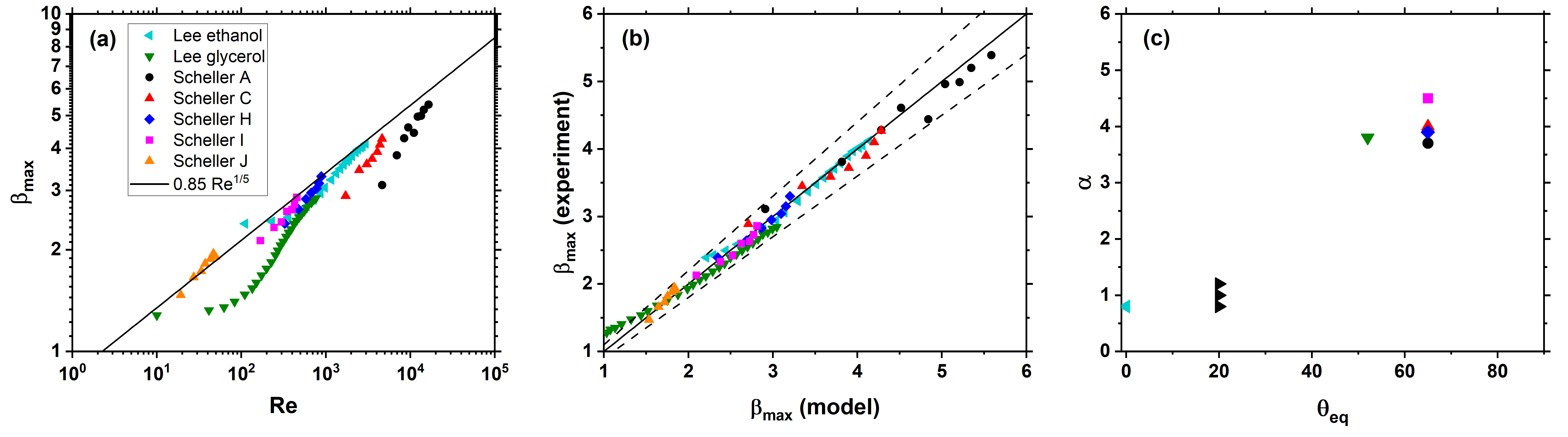}
  \caption{Literature data on spreading of Newtonian fluid droplets from Lee \textit{et al.}\cite{Lee2015} and Scheller \textit{et al.}\cite{scheller1993, Scheller1995}. Fluid parameters and corresponding fit parameter $\alpha$ are listed in Table \ref{tbl:lit parameters} (a) Spreading ratio versus Reynolds number. Solid line corresponds to viscous limit of our model [Eq. (\ref{Eqn:visclimit})]. (b) Comparison of model fit (Eq.(\ref{Eqn:EBM})) to literature data. (c) Fit parameter $\alpha$ versus equilibrium contact angle $\theta_{eq}$. Same symbols used as in (a) including our GO suspension data ($\blacktriangleright$).}
  \label{fgr:EBM Lit}
\end{figure*}
\begin{table*}[t]
\small
  \caption{\ Physical properties of Newtonian fluid droplet impact literature data with corresponding fit parameter $\alpha$ used in Fig.(\ref{fgr:EBM Lit}). For the highly viscous Scheller J fluid, the value of $\alpha$ has negligible influence on the fit and the viscous limit [Eq.(\ref{Eqn:visclimit})] was used.}
  \label{tbl:lit parameters}
  \begin{ruledtabular}
  \begin{tabular*}{\textwidth}{@{\extracolsep{\fill}}lllllllllll}
    Source & Sample & Substrate & Viscosity & Surface tension & Density & Radius& $\theta_{eq}$ & $\theta_d$ & $\theta_{\beta_{max}}$ & $\alpha$ fit \\
    & & & (mPa s)& (mN$/$m)& (kg$/{\text{m}}^3$)&mm & (deg.)& (deg.)& (deg.)&  \\
    \hline
    Scheller \textit{et al.} (1995)\cite{Scheller1995,scheller1993} & Water A &  Polystyrene & 1.0 & 72.0 & 992 & 1.8& 65 & -& -& 3.7 \\
     & Glycerol-water C &Polystyrene& 2.9 & 71.5 & 1082 & 1.8& 65& -& -& 4.0 \\
     & Glycerol-water H &Polystyrene& 16 & 68.5 & 1169 & 1.7& 65& -& -& 3.9\\
     & Glycerol-water I &Polystyrene& 31 & 68.0 & 1190 & 1.6& 65& -& -& 4.5\\
     & Glycerol-water J &Polystyrene& 300 & 65.0 & 1236 & 1.6& 65& -& -& Eq.(\ref{Eqn:visclimit})\\
    Lee \textit{et al.} (2015)\cite{Lee2015} & Ethanol & Steel & 1.2 & 23 & 789 & 0.9& 0& $\sim$44-90 & 44 & 0.8 \\
    & Glycerol-water & Steel & 10 & 68 & 1158 & 0.9& 52& $\sim$120-130 & 121 & 3.8 \\
  \end{tabular*}
  \end{ruledtabular}
\end{table*}
\subsection{Comparison of energy balance model with literature results}
We also fitted our model to literature data from Lee \textit{et al.}\cite{Lee2015} and Scheller \textit{et al.}\cite{scheller1993,Scheller1995} for Newtonian fluids as shown in Fig. \ref{fgr:EBM Lit} with fitted $\alpha$ values and the corresponding fluid parameters listed in table \ref{tbl:lit parameters}.  Fig. \ref{fgr:EBM Lit}(a) shows that $\beta_{max}$ is bounded by the viscous limit [Eq.(\ref{Eqn:visclimit})]. The model shows good agreement with the data as seen in Fig. \ref{fgr:EBM Lit}(b) with most of the data within $10\%$ of the model prediction. 

While the dynamic contact angle values are not available for the data from Scheller, Lee \textit{et al.} did measure the contact angle during spreading which allows us to estimate $\alpha$. In case for ethanol, the contact angle at maximum spread $\theta_{\beta_{max}}=44^\circ$, while the dynamic contact angle decreases smoothly from $90^\circ$ to around $\theta_{\beta_{max}}$ during the spreading (see table \ref{tbl:lit parameters}). Using the value for $\theta_{\beta_{max}}$ to estimate $\epsilon$ via the spherical cap approximation [Eq.(\ref{Eqn:cap})] and taking the average value for the dynamic contact angle $\theta_d$ to be $\approx 67^\circ$, we obtain $\alpha\approx 0.77$ [Eq.(\ref{Eqn:alpha})] which is in excellent agreement with the fitted value of $0.8$. For the Glycerol-water data from Lee \textit{et al.}, we have $\theta_{\beta_{max}}=121^\circ$ and $\theta_d\approx 125^\circ$ (see table \ref{tbl:lit parameters}), which gives $\alpha\approx 4.7$ which is larger than the fitted value of $3.8$. However, as pointed out by the Lee \textit{et al.}, the droplet shape deviates from a spherical cap, which leads to an overestimate of the interfacial area and is the likely explanation for this difference. Moreover, the value of $\epsilon$ in the spherical cap approximation is very sensitive for angles $\gtrsim100^\circ$.

Since $\alpha$ depends on the wetting properties we compared it with the equilibrium contact angle $\theta_{eq}$ as shown in Fig. \ref{fgr:EBM Lit}(c) for both our and the Newtonian literature data. It clearly shows that $\alpha$ is increasing with $\theta_{eq}$. In our model larger values of $\alpha$ lead to a lower minimum value of $\beta_{max}$ for a given fluid. This finding is consistent with experimental results from Lin \textit{et al.}\cite{lin2018} who found that the minimum value of $\beta_{max}$ at low impact speeds is decreasing with larger $\theta_{eq}$. 

While our energy balance model uses a simplified estimate for the viscous dissipation compared to other recent models\cite{Pasandideh-Fard1996,Wildeman2016,aksoy2022}, it captures the data well with only one fit parameter $\alpha$ that encodes both the dynamic contact angle and shape parameter $\epsilon$. The value of $\alpha$ also appears to be consistent with experimental values of the droplet shape and dynamic contact angle. Finally, our model allows for a analytic solution in the Newtonian limit. 

A constant value of $\alpha$ gives an excellent prediction of the experimental data at a given GO concentration. This may be surprising at first since the dynamic contact angle on which the value of $\alpha$ depends on increases appreciably with impact speed as shown in Fig.\ref{fgr:CA} while the shape parameter $\epsilon$ decreases with impact speed. However, since surface tension contributions in the energy balance only play a role at low impact speeds, the change in $\theta_d$ and $\epsilon$ does not affect the spreading ratio at high impact speeds where viscous dissipation dominates.

An open question is to what extent the value of $\alpha$, which depends on the dynamic contact angle and also sets the minimum value of $\beta_{max}$ [Eq.\ref{Eqn:EBMSol}] for a given fluid, can be inferred from the equilibrium contact angle $\theta_{eq}$. In general, the dynamic contact angle is between $\theta_{eq}$ and $180^\circ$ and increases with the capillary number though the flow field near the contact line also play a role\cite{sikalo2005,vadillo2009}. While Fig.\ref{fgr:EBM Lit}(c) suggests a correlation between $\alpha$ and $\theta_{eq}$ more data is needed to address to what extent other fluid parameters such as viscosity and droplet size influence the value of $\alpha$.

\section{Conclusions}
We have measured the spreading ratio of aqueous graphene oxide suspensions over a wide range of concentrations whose rheological response changes from Newtonian fluid to a shear-thinning, yield stress fluid. Using a characteristic viscosity the fluid experiences during the spreading, we formulated an energy balance model that incorporates both viscous dissipation and surface tension forces and can be solved analytically for Newtonian fluids. The model predictions are within $14\%$ of the experimental results. 
We have a single, non-dimensional fit parameter $\alpha$ in our model that encodes the droplet shape at maximum spread and the dynamic contact angle during spreading, both of which are not known a priori. The $\alpha$ parameter, which determines the minimum value of $\beta_{max}$, appears to be correlated with the equilibrium contact angle which opens up the possibility to predict the spreading ratio from the fluid and equilibrium wetting properties. 
\section*{Author Contributions}
MEM formulated the project and supervised the work. JAQ carried out the experiments and analysis. All authors discussed the results and wrote the paper.
\section*{Acknowledgements}
The authors would like to thank S. Barwich for help with the sample preparation and GO particle characterisation. MEM would like to thank Prof. Martin Hegner for giving us access to the MD-P-801 autodrop system. The research was supported by the grant SFI 17/CDA/4704 and SFI AMBER2 12/RC/2278. This article is distributed under a Creative Commons Attribution-NonCommercial-NoDerivs 4.0 International (CC BY-NC-ND) License.
\section*{Data Availability Statement}
The data that support the findings of this study are available from the corresponding author upon reasonable request.
\bibliography{Spreadingpaper.bib}

\end{document}